# *Option Pricing: Channels, Target Zones and Sideways Markets*

Zura Kakushadze[§†1]

§ *Quantigic® Solutions LLC,[2] 680 E Main St, #543, Stamford, CT 06901*
† *Free University of Tbilisi, Business School & School of Physics*
*240, David Agmashenebeli Alley, Tbilisi, 0159, Georgia*

May 25, 2020

Abstract

After a market downturn, especially in an uncertain economic environment such as the current state, there can be a relatively long period with a sideways market, where indexes, stocks, etc., move in channels with support and resistance levels. We discuss option pricing in such scenarios, in both cases of unattainable as well as attainable boundaries, and obtain closed-form option pricing formulas. Our results also apply to FX rates in target zones without interest rate pegging (USD/HKD, digital currencies, etc.).

---

[1] Zura Kakushadze, Ph.D., is the President and a Co-Founder of Quantigic® Solutions LLC and a Full Professor in the Business School and the School of Physics at Free University of Tbilisi. Email: zura@quantigic.com

[2] DISCLAIMER: This address is used by the corresponding author for no purpose other than to indicate his professional affiliation as is customary in publications. In particular, the contents of this paper are not intended as an investment, legal, tax or any other such advice, and in no way represent views of Quantigic® Solutions LLC, the website www.quantigic.com or any of their other affiliates.

## 1. Introduction and Summary

After a market downturn, especially in an uncertain economic environment such as the current state, it would not be surprising to have a relatively long period with a sideways market, where indexes and many individual securities trade in relatively tight ranges, without a clear up or down trend. Thus, a stock (or some other security) can be moving in a channel with support and resistance levels. So, we ask: *How should we price options on a stock moving in a channel?*

This is not a rhetorical question. Thus, if one had "divine" knowledge that the market would be going sideways for some period of time, buying calls or puts would make little sense. Instead, one would be happily selling volatility. By the same token, it would make little sense to assume the usual distributions (with or without fat tails) for stock returns when pricing options.

One way to approach the problem is to assume that stock returns are mean-reverting (see, e.g., [Madan, 2017]), without assuming that the prices are confined to a channel. This is a valid approach; however, things get complicated quickly, and analytic tractability is infeasible.

To achieve analytic tractability, here we take a different approach. We assume that a stock is actually moving between well-defined upper and lower boundaries. Two conceptually different cases arise in this context. We can have unattainable boundaries, i.e., the stock price never touches the boundaries, only approaching them asymptotically. We discuss this case in Section 2. One alluring feature of unattainable boundaries is that bounded martingales can be constructed relatively easily (assuming zero risk-free rate, which is not farfetched in the current environment; a nonzero risk-free rate makes things more involved). In fact, in the simple model of [Kakushadze, 2019] we can compute option prices analytically. On the flipside, such models are mean-repelling, so in the long term the stock price ends up close to one of the boundaries.

This is avoided when boundaries are attainable, which are also more in line with the real life. So that probability does not leak through them, the boundaries must be reflecting. Then it is unavoidable to have arbitrage at the boundaries, so arbitrage-free option pricing via self-financing hedging strategies is no longer possible. Happily, as we discuss in Section 3, precisely due to arbitrage, the call (put) price must vanish at the lower (upper) boundary, and the pricing problem reduces to that of double-barrier knockout options with rebates (immediately payable upon hitting the upper (lower) boundary, which can be recast in terms of a cap with automatic exercise), which is well-understood, including for a nonzero risk-free rate, with dividends, etc.

Another application of our results is to managed FX rates in target zones (e.g., USD/HKD or digital currencies). Option pricing where the interest rate is pegged (so there is no arbitrage) is discussed in [Carr and Kakushadze, 2017]. As discussed in [Kakushadze and Yu, 2019], such a peg cannot be achieved in realistic high-volatility scenarios. So our results can be used instead.

## 2. Unattainable Barriers

Let the stock (or, more generally, some instrument) price $S_t$ be modeled via

$$S_t = \exp(X_t) \quad (1)$$

$$dX_t = \mu(X_t)\, dt + \sigma\, dW_t \quad (2)$$

The drift $\mu(X_t)$ has no explicit dependence on time $t$, and the volatility $\sigma$ is constant.[3] $W_t$ is a **P**-Brownian motion (a.k.a. a Wiener process), with null drift and variance $t$ (**P** is the measure).

Let $M_t = w(X_t, t)$ be a **P**-martingale, where $w(x, t)$ is a deterministic function. Since $M_t$ is driftless, $w(x, t)$ must satisfy the following partial differential equation (PDE):

$$\partial_t w(x,t) + \mu(x)\, \partial_x w(x,t) + \frac{\sigma^2}{2}\, \partial_x^2 w(x,t) = 0 \quad (3)$$

This equation admits "zero-mode" solutions with no explicit time dependence: $\partial_t w(x,t) \equiv 0$. Thus, we have ($C \neq 0$ is an integration constant):

$$\partial_x w(x) = C \exp\left(-\frac{2}{\sigma^2} \int_0^x \mu(y)\, dy\right) \quad (4)$$

If $\mu(x)$ goes to a positive (negative) constant or to the positive (negative) infinity as $x$ goes to the positive (negative) infinity, then $\partial_x w(x)$ vanishes at the positive (negative) infinity, so $w(x)$ is bounded there. So, for such $\mu(x)$ we have a martingale $M_t = w(X_t)$ bounded between finite values $w_\pm = w(\pm\infty)$. The boundaries $w_\pm$ are *unattainable*. Furthermore, the process $X_t$ is *mean-repelling*. An example of such a process (with linear $\mu(x)$) was discussed in [Carr, 2017].

There is a choice of $\mu(x)$ for which both the bounded martingale $M_t = w(X_t)$ and the probability density are particularly simple allowing for analytically tractable option pricing. This model was discussed in [Kakushadze, 2019]:

$$\mu(x) = \nu\, \sigma^2 \tanh(\nu\, (x - x_*)) \quad (5)$$

Here $\nu$ is a constant. The bounded martingale then reads ($A$ and $B$ are integration constants):

$$w(x) = A + B \tanh(\nu\, (x - x_*)) \quad (6)$$

In this case we have the (unattainable) boundaries $w_\pm = A \pm B$ (and the "center" is at $x = x_*$).

Let $P(x, x_0; t, 0)$ be the probability distribution of starting at $X_t = x_0$ at time $t = 0$ and ending at $X_t = x$ at time $t$. The Fokker-Planck equation reads:

---
[3] We could be more general and consider the volatility dependent on the state variable $X_t$. We will not do so here.

$$\partial_t P + \partial_x[\mu(x)\, P] - \frac{\sigma^2}{2}\, \partial_x^2 P = 0 \qquad (7)$$

The solution for the probability density (normalized to 1 when integrated over $x$ from $-\infty$ to $+\infty$) is given by [Kakushadze, 2019]:

$$P(x, x_0; t, 0) = \frac{1}{\sqrt{2\pi t}\, \sigma}\, \frac{\cosh(\nu\,(x - x_*))}{\cosh(\nu\,(x_0 - x_*))}\, \exp\left(-\frac{(x - x_0)^2}{2\sigma^2 t} - \frac{\sigma^2 \nu^2 t}{2}\right) \qquad (8)$$

So, the probability density, conveniently, is a linear combination of two Gaussians in this model.

Before we discuss option pricing, let us note that asymptotically, as $X_t \to \pm\infty$, the stock process $S_t$ becomes a geometric Brownian motion with a positive/negative drift. So, we expect martingales other than the bounded martingale to exist. These other martingales differ from the time-homogeneous bounded martingale (Eq. (6)) in that they have explicit $t$-dependence. Time-inhomogeneous martingales solving Eq. (3) can be constructed by separating variables in a standard fashion. Thus, let

$$w(x, t) = \frac{\phi(x)}{\cosh(\nu\,(x - x_*))}\, \exp\left(-\frac{\rho\, \sigma^2 t}{2}\right) \qquad (9)$$

Here $\rho$ is a constant parameter, and the function $\phi(x)$, to be determined, is independent of $t$. Substituting $w(x, t)$ given by Eq. (9) into Eq. (3), we have the following equation for $\phi(x)$:

$$\partial_x^2 \phi(x) - [\rho + \nu^2]\, \phi(x) = 0 \qquad (10)$$

The two independent solutions are given by $\phi(x) = \exp(\pm\lambda\, x)$, where $\lambda = \sqrt{\rho + \sigma^2}$. The corresponding martingales are unbounded and monotonic functions of $x$ (assuming $\rho > 0$).[4]

## 2.1. Option Pricing

Eq. (1) assumes that the stock process is exponential in the state variable $X_t$. However, this assumption was redundant in that it played no role in deriving the probability density (Eq. (8)) or the bounded martingale (Eq. (6)). Thus, there is nothing stopping us from assuming that the measure **P** is the *risk-neutral* measure, and that the *discounted* stock process[5] $Z_t = \exp(-rt)\, S_t$ is given by $M_t = w(X_t)$, where $w(x)$ is given by Eq. (6). Here we are assuming a constant risk-free interest rate $r$.[6] Here we need to deal with two issues. The first is that the stock process $S_t$ must be positive, which is ensured by having positive boundaries $w_\pm$, i.e., $A > B > 0$ in Eq. (6). The second issue is that, unless the risk-free rate $r = 0$, the boundaries for the stock process $S_t$ are not static. In fact, this statement holds not only for the specific drift given by Eq. (5), but for a *generic* drift. Thus, assuming a nonzero risk-free rate, to have static boundaries for the stock process $S_t = \exp(rt)\, w(X_t, t)$, we have $w(x, t) = \exp(-rt)\, \varphi(x)$ and the following differential equation for $\varphi(x)$:

$$-r\, \varphi(x) + \mu(x)\, \partial_x \varphi(x) + \frac{\sigma^2}{2}\, \partial_x^2 \varphi(x) = 0 \tag{11}$$

---

[4] For $\rho < 0$ martingales are bounded, but are not monotonic functions of $x$ and are not suitable for option pricing.
[5] For details, see, e.g., [Baxter and Rennie, 1996], [Harrison and Pliska, 1981], [Hull, 2012], [Kakushadze, 2015].
[6] This is not critical here and we can have a variable (and even nondeterministic) interest rate. What is of import is that the interest rate is not "pegged" to the stock price $S_t$ (and, if non-constant, is governed by its own dynamics).

Recall that $\varphi(x)$ is a positive monotonic function (see above) with unattainable boundaries at $x \to \pm\infty$, which implies that $\partial_x \varphi(x \to \pm\infty) = 0$ and $\partial_x^2 \varphi(x \to \pm\infty) = 0$. Then, assuming (without loss of generality) that $\varphi(x)$ is a monotonically increasing function, Eq. (11) implies that $\mu(x \to \pm\infty) \to \text{sign}(r) \times \infty$. Furthermore, $|\mu(x)|$ must grow faster than $|x|$ at $x \to \pm\infty$ (or else $\varphi(x)$ will not be finite at $x \to \pm\infty$). We can view Eq. (11) as an equation for $\mu(x)$:

$$\mu(x) = r\, \frac{\varphi(x)}{\partial_x \varphi(x)} + \tilde{\mu}(x) \tag{12}$$

Here $\tilde{\mu}(x)$ is a (simpler) drift we may wish to choose in the $r = 0$ case (e.g., that given by Eq. (5)) and $\varphi(x)$ satisfies the following "zero-rate" equation

$$\tilde{\mu}(x)\, \partial_x \varphi(x) + \frac{\sigma^2}{2}\, \partial_x^2 \varphi(x) = 0 \tag{13}$$

Eq. (12) implies that $\mu(x)$ is not mean-repelling, but "mean-reverting" in one asymptotic regime and "mean-repelling" in the other (and, as mentioned above, blows up faster-than-linearly at infinity). Furthermore, it is not generic in the sense that it is "fine-tuned" for a given value of the risk-free rate. This is unsurprising. Indeed, for a nonzero risk-free rate the time value of money makes static boundaries unnatural and extra "gymnastics" are needed to achieve them.[7]

The price for such "gymnastics" is that things get complicated very quickly and analytic tractability is lost. On the other hand, for $r = 0$ (which is not farfetched at all currently), in the simple model given by Eq. (5), we can compute option prices analytically. The price $V_c(t, T, K)$ of a vanilla European call option with the strike price $K$ and maturity $T$ is given by

$$V_c(t, T, K) = \int_{-\infty}^{+\infty} dx\, P(x, x_0; T, t)\, (w(x) - K)^+ \tag{14}$$

Here: the probability density $P(x, x_0; T, t)$ is given by Eq. (8); $x_0$ corresponds to the stock price $S_t$ at time $t$, i.e., $S_t = A + B \tanh(v x_0)$ (see Eq. (6)), where without loss of generality we have set $x_* = 0$); and $(z)^+ = \max(z, 0)$. The integral in Eq. (14) is straightforward to compute as the probability density is a sum of two Gaussians:

$$V_c(t, T, K) = B \frac{\exp(v(x_0 - x_1)) N(d_+) - \exp(-v(x_0 - x_1)) N(d_-)}{2 \cosh(vx_0) \cosh(vx_1)} \quad (15)$$

---

[7] Unattainable static boundaries can be achieved by having the volatility dependent on $X_t$ and vanishing at the boundaries. In the context of FX rates in target zones, see, e.g., various references in [Carr and Kakushadze, 2017].

Here $x_1$ corresponds to the strike price $K$, i.e., $K = A + B \tanh(vx_1)$; $N(z) = \Phi(z)$ is the cumulative normal distribution function; and

$$d_\pm = \frac{x_0 - x_1}{\sigma\sqrt{T - t}} \pm v\sigma\sqrt{T - t} \quad (16)$$

The price $V_p(t, T, K)$ of a vanilla European put option with the strike price $K$ and maturity $T$ follows from the put-call parity and therefore is given by $V_p(t, T, K) = V_c(t, T, K) - S_t + K$.

Let us note that Eq. (15) can be neatly rewritten as follows:

$$V_c(t, T, K) = \frac{[S_t - S_-][S_+ - K] N(d_+) - [S_+ - S_t][K - S_-] N(d_-)}{S_+ - S_-} \quad (17)$$

Here $S_\pm = w_\pm = A \pm B$ are the upper and lower boundaries. So, when the time to maturity $T - t$ is large, we have

$$V_c(t, T, K) \approx \frac{[S_t - S_-][S_+ - K]}{S_+ - S_-} \quad (18)$$

This result has a simple underlying interpretation. As mentioned above, the probability density given by Eq. (8) is a sum of two Gaussians. It can be written as follows (again, we set $x_* = 0$):

$$P(x, x_0; T, t) = \frac{S_t - S_-}{S_+ - S_-} P_+(x, x_0; T, t) + \frac{S_+ - S_t}{S_+ - S_-} P_-(x, x_0; T, t) \quad (19)$$

$$P_\pm(x, x_0; T, t) = \frac{1}{\sqrt{2\pi(T - t)}\, \sigma} \exp\left(-\frac{(x - x_0 \mp \mu_*(T - t))^2}{2\sigma^2(T - t)}\right) \quad (20)$$

Here $\mu_* = \sigma^2 v$. So, for large $T - t$ the stock price is pushed either to the upper boundary with probability $\tilde{P}_+ = (S_t - S_-)/(S_+ - S_-)$, or to the lower boundary with probability $\tilde{P}_- = (S_+ - S_t)/(S_+ - S_-)$. Hence the result given by Eq. (18), which is nothing but the probability $\tilde{P}_+$ of being pushed to the upper boundary times the payoff, which is $S_+ - K$. The underlying reason for this behavior is that we have a mean-repelling process, so in the long term the stock process always ends up being pushed to either boundary. Therefore, the above approach is somewhat removed from reality, as are unattainable boundaries. Also, to have static boundaries and keep analytic tractability, we had to set the risk-free rate to zero. On the other hand, if we employ the volatility dependent on $X_t$, then things get complicated quickly w.r.t. analytic tractability.

## 3. Attainable Boundaries

As we saw above, unattainable boundaries are no free lunch. So, let us switch gears and consider attainable boundaries, i.e., the stock process $S_t$ moves between the upper $S_+$ and lower $S_-$ boundaries and $S_t$ from time to time can and does take these boundary values. With attainable boundaries we invariably must impose boundary conditions. In this context, the boundary conditions are in terms of the underlying state variable $X_t$. Thus, as above, we will assume that the stock process is some function of the state variable: $S_t = f(X_t, t)$. If we wish to have static boundaries, then we must have no explicit time-dependence: $S_t = f(X_t)$. And, as above, we will also assume that the state variable stochastic dynamics is modeled via Eq. (2).

A priori, we could impose reflecting (Neumann), absorbing (Dirichlet) or mixed (Robin) boundary conditions. The boundary conditions determine the probability density, martingales, etc. The proper choice is to have reflecting boundary conditions. The reason for this is simple (see, e.g., [Carr and Kakushadze, 2017]). Unless we have reflecting boundary conditions, the probability will leak through the boundaries. A more intuitive way of stating the same is that the identity process $I_t \equiv 1$ must be a martingale (or else the integral of the probability density $P(x, x_0; T, t)$ (using the same notations as above) over $x$ between $x_-$ and $x_+$ (corresponding to $S_-$ and $S_+$, respectively) will not equal 1. However, the price of this is that, while $I_t$, as it should be, is a martingale, for a nonzero risk-free rate we cannot construct another martingale that would correspond to the discounted stock price. Such a martingale $M_t = w(X_t) \exp(-rt)$ must satisfy the conditions that $w(x)$ is a monotonic positive function of $x$ with the reflecting boundary conditions $\partial_x w(x_\pm) = 0$. However, this is incompatible with the martingale PDE, which reduces to

$$-r\, w(x) + \mu(x)\, \partial_x w(x) + \frac{\sigma^2}{2}\, \partial_x^2 w(x) = 0 \qquad (21)$$

Indeed, we invariably have that $\partial_x^2 w(x_\pm)$ are both positive (for $r > 0$) or negative (for $r < 0$), so at both $x_\pm$ we have minima (for $\partial_x^2 w(x_\pm) > 0$) or maxima (for $\partial_x^2 w(x_\pm) < 0$), incompatibly with the monotonicity of $w(x)$. Furthermore, for $r \neq 0$ the boundaries would not be static.[8]

The issue here is that in the presence of reflecting boundaries arbitrage is unavoidable, so the standard arbitrage-free pricing arguments involving self-financing hedging strategies (see, e.g., [Baxter and Rennie, 1996], [Carr and Kakushadze, 2017], [Harrison and Pliska, 1981], [Hull, 2012], [Kakushadze, 2015]) no longer go through. It is simple to see this when the risk-free rate vanishes. Then, if the stock process ever hits the lower (upper) boundary, since the boundaries are reflective, we are guaranteed risk-free profits by buying (selling) the stock at that boundary. When the risk-free rate is positive, then it is less evident that borrowing $S_-$ dollars and buying at the lower boundary yields a risk-free profit. Indeed, if we simply hold the stock for some finite time $t$ following the purchase at $t = 0$, its price $S_t$ may be below the amount we owe at that time, which is $S_- \exp(rt)$. However, there is still arbitrage. A quick, intuitive way to see this in the continuous time setting is to note that in the infinitesimal time $dt$ the stock price moves away from the lower boundary by an order of magnitude of $\sigma\sqrt{dt}$ (on a relative basis), while the debt increases by an order of magnitude of $rdt$, so the stock "wins".

---

[8] If the risk-free rate vanishes, for contrived drifts we can construct formal solutions to Eq. (21) that are positive monotonic functions of $x$ with the reflecting boundary conditions. Such $w(x)$ have inflection points at $x_-$ and $x_+$. For such a function to have a continuous second derivative, the latter must also vanish in the interior of $[x_-, x_+]$. However, this does not mean that we can construct a self-financing hedging strategy. Mathematically, this is due to the fact that $\partial_x^2 w(x)$ vanishes at the boundaries, which makes Delta-hedging at the boundaries impossible (see, e.g., Subsection 3.2 of [Carr and Kakushadze, 2017]). Financially, reflecting boundaries imply arbitrage (see below).

We can make this more precise by considering a discrete binary tree instead. First, let us ignore the boundaries and consider pricing claims on a binary tree in the interior, away from the boundaries. At time $t$ the stock price is $S_{now}$. At time $t + \delta t$ the stock price can take two values: $S_{up}$ and $S_{down}$. At time $t$ the bond is worth $B_{now}$, and at time $t + \delta t$ the bond value is $B_{now} \exp(r\delta t)$. Further, suppose we have a claim $f$, which at time $t + \delta t$ takes two values $f_{up}$ and $f_{down}$ according to the stock price. We can synthesize this derivative using a self-financing hedging strategy (see the references cited above). So, the worth of the claim at time $t$ is given by $f_{now} = \exp(-r\delta t)\,[q\, f_{up} + (1-q)\, f_{down}]$, where $q = [\exp(r\delta t)\, S_{now} - S_{down}]\,/\,[S_{up} - S_{down}]$. The set $\mathbf{Q} = \{q, 1-q\}$ is the risk-neutral measure. The fact that $q > 0$ follows from the fact that otherwise we would have $\exp(r\delta t)\, S_{now} \leq S_{down} < S_{up}$, which would guarantee unlimited risk-free profit at time $t + \delta t$ by selling the cash bond and buying stock at time $t$. We also have $q < 1$, which follows from the fact that otherwise we would have $\exp(r\delta t)\, S_{now} \geq S_{up} > S_{down}$, which would guarantee unlimited risk-free profit at time $t + \delta t$ by selling the stock and buying the cash bond at time $t$. So, we must have $S_{down} < \exp(r\delta t)\, S_{now} < S_{up}$.

Now consider $S_{now}$ at time $t$ being at the lower boundary: $S_{now} = S_-$. Since the boundary is reflecting, the stock price cannot go down, it can only go up to $S_{up}$. It then follows that there is arbitrage as borrowing $S_{now}$ dollars and buying the stock at time $t$ guarantees risk-free profit $S_{up} - \exp(r\delta t)\, S_{now} > 0$. A similar argument also applies to the upper boundary.

### *3.1. Rational Option Pricing?*

So, with reflecting boundaries we always have arbitrage. How can we price claims such as European call and put options in the presence of reflecting boundaries then? Despite the presence of arbitrage, two rational agents should be able to agree on a "fair" price for a claim.

Here we can think about this problem from a purely pragmatic viewpoint. Consider a European call option with the strike price $K$ and maturity $T$ (here $S_- < K < S_+$). Let us assume that at time $t$ the stock price $S_- < S_t < S_+$. We can ask, what happens if the stock price ever hits $S_-$? The answer is that the option should become worthless. Indeed, at $S_-$ we have arbitrage, so there is no point in paying for any claim. We can also ask, what happens if the stock price ever hits $S_+$? The answer is that the call option should be automatically exercised as at $S_+$ we also have arbitrage. So, our call option has the following boundary condition: $V_c(t,T,K) = 0$ if $S_t = S_-$, and it is automatically exercised if $S_t = S_+$. That is, we have a knockout at the lower boundary. In fact, we can view this claim as a *double-barrier knockout* call option, both at the lower and upper boundaries (i.e., an up-and-out-down-and-out call option), with the *rebate* equal $S_+ - K$ at the upper boundary. So, the price of our call option is

$$V_c(t,T,K) = V_c^{dko}(t,T,K,S_-,S_+) + (S_+ - K)\, V_b^{upper}(t,T,S_-,S_+) \qquad (22)$$

Here $V_c^{dko}(t,T,K,S_-,S_+)$ is the price of the European double-barrier knockout call option, which pays $(S_T - K)^+$ at maturity $T$ if $S_- < S_t < S_+$ before maturity, or else (i.e., if either of the boundaries is touched) it pays 0. On the other hand, $V_b^{upper}(t,T,S_-,S_+)$ is the price of an *American double-barrier binary asymmetrical* option, which, at any time before maturity $T$, is knocked out if the stock price hits the lower boundary, and immediately pays \$1 if the stock price hits the upper boundary. If the stock dynamics is governed by a geometric Brownian motion, the call option price $V_c^{dko}(t,T,K,S_-,S_+)$ is given in [Kunitomo and Ikeda, 1992],[9] while the binary option price $V_b^{upper}(t,T,S_-,S_+)$ is given in [Hui, 1996]; both are also in [Haug, 2007].

A European put option can be priced similarly. It has the following boundary condition: $V_p(t,T,K) = 0$ if $S_t = S_+$, and it is automatically exercised if $S_t = S_-$. Our put option price is

$$V_p(t,T,K) = V_p^{dko}(t,T,K,S_-,S_+) + (K - S_-)\, V_b^{lower}(t,T,S_-,S_+) \qquad (23)$$

Here $V_p^{dko}(t,T,K,S_-,S_+)$ is the price of the European double-barrier knockout put option (i.e., an up-and-out-down-and-out put option), which pays $(K - S_T)^+$ at maturity $T$ if $S_- < S_t < S_+$ before maturity, or else (i.e., if either of the boundaries is touched) it pays 0. Furthermore, $V_b^{lower}(t,T,S_-,S_+)$ is the price of an American double-barrier binary asymmetrical option, which, at any time before maturity $T$, is knocked out if the stock price hits the upper barrier, and immediately pays \$1 if the stock price hits the lower barrier. (See the references above.)

For the call (but not put) option the result of Eq. (22) can be extended to American options (the call option can be exercised at any time before maturity) with no dividends as early exercise is suboptimal. Dividends can introduce an incentive to exercise early. If dividends are sufficiently low, the optimal exercise policy remains to exercise at the upper boundary, so the American exercise feature is worthless; see, e.g., [Broadie and Detemple, 1995] (and references therein) for the pricing of capped American calls with low dividends. In this regard, let us note that the upper boundary plays the same role as the cap in a capped call option. Thus, Eqs. (22) and (23) are not affected by reflecting boundary conditions as the Brownian motion is *local* and the options are knocked out or exercised at the boundaries, so their reflecting property is moot.

---

[9] For other works, also see [Beaglehole, 1992], [Bhagavatula and Carr, 1995], [Carr, 1995], [Geman and Yor, 1996].